\documentstyle[epsf]{paper}
\def\NH      {N_{\rm H}}
\def\Rwd     {R_{\rm wd}}
\def\Mwd     {M_{\rm wd}}
\def\Msun    {{\rm M_{\odot}}}
\def\Mdot    {\mathaccent 95 M}
\def\peryr   {{\rm yr^{-1}}}
\newbox\grsign \setbox\grsign=\hbox{$>$} 
\newdimen\grdimen \grdimen=\ht\grsign
\newbox\laxbox \newbox\gaxbox
\setbox\gaxbox=\hbox{\raise.5ex\hbox{$>$}\llap
     {\lower.5ex\hbox{$\sim$}}}\ht1=\grdimen\dp1=0pt
\setbox\laxbox=\hbox{\raise.5ex\hbox{$<$}\llap
     {\lower.5ex\hbox{$\sim$}}}\ht2=\grdimen\dp2=0pt
\def\gax{\mathrel{\copy\gaxbox}}
\def\lax{\mathrel{\copy\laxbox}}
\newcounter{nmbr}
\setbox0=\hbox{\rm00.~}
\newdimen\widtha
\widtha=\wd0
\setbox1=\hbox{\rm00}
\newdimen\widthb
\widthb=\wd1
\newcommand{\reference}{\addtocounter{nmbr}{1}
                        \par\hangindent=\widtha\hangafter=1\noindent
                        \hbox to \widthb{\hfil\thenmbr}.~}

\begin{document}

\title{X-RAY AND EUV SPECTROSCOPY OF THE BOUNDARY LAYER EMISSION OF
NONMAGNETIC CATACLYSMIC VARIABLES}

\author{Christopher W.\ Mauche \\
{\it Lawrence Livermore National Laboratory,\\
L-41, P.O.\ Box 808, Livermore, CA 94550, USA}}

\maketitle

\vbox to 0pt{\vskip -6.0cm
\hbox to \hsize{%
{\it 1997, in X-ray Imaging and Spectroscopy of Cosmic Hot Plasmas,\hfil }}
\hbox to \hsize{%
{\it ed.~F.~Makino \& K.~Mitsuda (Tokyo: Universal Academy Press), 529\hfil }}
\vss}
\vspace{-12pt}

\section*{Abstract}

{\it EUVE\/}, {\it ROSAT\/}, and {\it ASCA\/} observations of the boundary
layer emission of nonmagnetic cataclysmic variables (CVs) are reviewed. {\it
EUVE\/} spectra reveal that the effective temperature of the soft component of
high-$\Mdot $ nonmagnetic CVs is $kT\sim 10$--20 eV and that its luminosity is
$\sim 0.1$--0.5 times the accretion disk luminosity. Although the EUV spectra
are very complex and belie simple interpretation, the physical conditions
of the boundary layer gas are constrained by emission lines of highly ionized
Ne, Mg, Si, and Fe. {\it ROSAT\/} and {\it ASCA\/} spectra of the hard
component of nonmagnetic CVs are satisfactorily but only phenomenologically
described by multi-temperature thermal plasmas, and the constraints imposed
on the physical conditions of this gas are limited by the relatively weak and
blended lines. It is argued that significant progress in our understanding of
the X-ray spectra of nonmagnetic CVs will come with future observations with
{\it XMM\/}, {\it AXAF\/}, and {\it Astro-E\/}.

\section{Introduction}

In the standard theory of disk accretion in nonmagnetic cataclysmic variables,
half of the gravitational potential energy of accreted material is dissipated
in the disk, and half is dissipated in the boundary layer between the disk
and the surface of the white dwarf. The luminosity of the boundary layer is
$L=G\Mwd\Mdot/2\Rwd= 8\times 10^{34}\, (\Mwd/\Msun )(\Mdot/10^{-8}\,
\Msun\peryr)\allowbreak (\Rwd/5\times 10^8~{\rm cm})^{-1}~\rm erg~s^{-1}$,
where $\Mwd $ is the mass of the white dwarf, $\Rwd $ is its radius, and
$\Mdot $ is the mass-accretion rate. The temperature of the boundary layer
is bounded from below by the blackbody temperature $kT_{\rm low}=
k(G\Mwd\Mdot/8\pi\sigma\Rwd ^3)^{1/4}\sim 10$ eV and from above by the virial
temperature $kT_{\rm high} =G\Mwd\mu m_{\rm H}/3\Rwd \sim 60$ keV. The former
applies in high-$\Mdot $ systems (i.e., nova-like variables and dwarf novae in
outburst) wherein the boundary layer is optically thick and the  radiation is
thermalized before escaping, and the latter applies in low-$\Mdot $ systems
(i.e., dwarf novae in quiescence) wherein the boundary layer is optically
thin. However, even high-$\Mdot $ systems have a high-energy component in
their spectra due to the inevitable optically thin portion of the boundary
layer. The physical conditions in the boundary layer are determined by
radiation hydrodynamic processes which are too complex to model with
fidelity, and depend on such unknowns as the rotation rate of the white dwarf
and the nature and magnitude of the viscosity. Insight into the physical
conditions in the boundary layer and its environs is afforded by 
spectroscopic studies in the extreme ultraviolet (EUV) through hard X-rays.

\section{ROSAT Observations}

{\it ROSAT\/} measured the X-ray spectra of CVs in an intermediate bandpass
($E=0.1$--2.4 keV) with moderate spectral resolution ($E/\Delta E\sim 2.4$ at
1~keV). The broad picture of the X-ray spectra of $\sim 120 $ CVs is supplied
by the {\it ROSAT\/} All-Sky Survey (Beuermann \& Thomas 1993). Insufficient
counts are typically collected to perform spectral fits, but useful
diagnostics are provided by such hardness ratio diagrams as the total count
rate versus the hardness ratio $\rm (hard-soft)/(hard+soft)$, where ``soft''
is the 0.07--0.40 keV flux and ``hard'' is the 0.40--2.40 keV flux. 
High-$\Mdot $ AM Her-type CVs have hardness ratios $<-0.5$ and, with the
exception of SS~Cyg in outburst, both low- and high-$\Mdot $ nonmagnetic CVs
have hardness ratios $>-0.5$. The soft spectra of high-$\Mdot $ AM Her-type
CVs is due to the expression of a large fraction of the accretion luminosity
in a soft blackbody-like component having a temperature of a few tens of eV.
The moderate-to-hard ratios of nonmagnetic CVs can be reproduced by a thermal
brems spectrum with a temperature of a few to many keV. Absorption by a
neutral hydrogen column drastically reduces the total count rate with only a
modest variation in the hardness ratio for soft blackbody spectra, but under
the same influence thermal brems spectra suffer large variations in the
hardness ratio with only modest variations in the total count rate. Despite
the many high-$\Mdot $ nonmagnetic CVs observed during the All-Sky Survey,
only SS~Cyg in outburst clearly manifests a soft component in its spectrum
(Ponman et~al.\ 1995). When VW~Hyi went into outburst, its hardness ratio
remained constant while the total count rate {\it decreased\/} (Wheatley
et~al.\ 1996). The soft component of high-$\Mdot $ nonmagnetic CVs is not
detected because either its luminosity is significantly smaller than the
high-temperature component, its temperature is too low to radiate soft X-rays,
its flux is extinguished by photoelectric absorption in the ISM or a wind, or
some combination of these factors. To distinguish among these alternatives,
{\it EUVE\/} observations are required (\S 3).

Analysis of pointed {\it ROSAT\/} observations of nonmagnetic CVs is supplied
by van Teeseling \& Verbunt (1994), van Teeseling, Beuermann, \& Verbunt
(1996), Richman (1996), and references therein. The colors of nonmagnetic CVs
can be reproduced by absorbed single-temperature thermal brems or optically
thin thermal plasma spectra, but fits of the data to such models demonstrate
that the spectra of nonmagnetic CVs are often more complex. van Teeseling \&
Verbunt found that acceptable fits could be obtained with either absorbed
two-temperature optically thin thermal plasma spectra or absorbed
single-temperature thermal brems spectra with a narrow emission line at $\sim
1$ keV. Richman found that fits to absorbed thermal brems spectra produced
excesses near 0.2 and 0.9 keV. Two-temperature absorbed thermal brems spectra
did not significantly improve the fits, and although the addition of a narrow
emission line at $\sim 1$ keV did improve the fits, most of the reduction in
$\chi ^2$ was due to the additional parameters in the model. Unfortunately,
although {\it ROSAT\/} PSPC data is an improvement over {\it Einstein\/} IPC
data in being able to tell that the spectra of nonmagnetic CVs are more
complex than absorbed single-temperature models, the PSPC has neither the
bandpass nor the spectral resolution required to distinguish between the
alternatives. As such, the increase in complexity has not been accompanied by
an increase in understanding. To accomplish this, the wider bandpass and
higher spectral resolution of {\it ASCA\/} (\S 4), {\it XMM\/}, {\it AXAF\/},
and {\it Astro-E\/} are required (\S 5).

\section{EUVE Observations}

Observations in the EUV are needed to measure the soft component of the X-ray
spectra of high-$\Mdot $ CVs, but photoelectric absorption severely limits
the effective bandpass and the number of sources that can be observed. Unit
optical depth is reached for a hydrogen column density of $\NH =10^{18}$,
$10^{18.5}$, $10^{19}$, $10^{19.5}$, and $10^{20}~\rm cm^{-2}$ at $\lambda
\sim 380$, 235, 145, 95, and 65~\AA , respectively. With the exception of
VW~Hyi, whose interstellar column density is $\sim 6\times 10^{17}~\rm
cm^{-2}$, CVs have interstellar columns in excess of a couple times
$10^{19}~\rm cm^{-2}$. To make matters worse, high-$\Mdot $ nonmagnetic CVs
have winds which can absorb and scatter the boundary layer radiation. {\it
EUVE\/} supplies high-resolution ($\lambda/\Delta\lambda \sim 200$) spectra
in the bandpass from 70 to 760~\AA , but observations exist for only a few
high-$\Mdot $ nonmagnetic CVs: the nova-like variable IX~Vel (van Teeseling
et~al.\ 1995) and the dwarf novae U~Gem (\S 3.1), SS~Cyg (\S 3.2), and VW~Hyi
(\S 3.3) in outburst.

\newpage

\begin{figure}[h] 
\epsfbox[ 86 545 431 700]{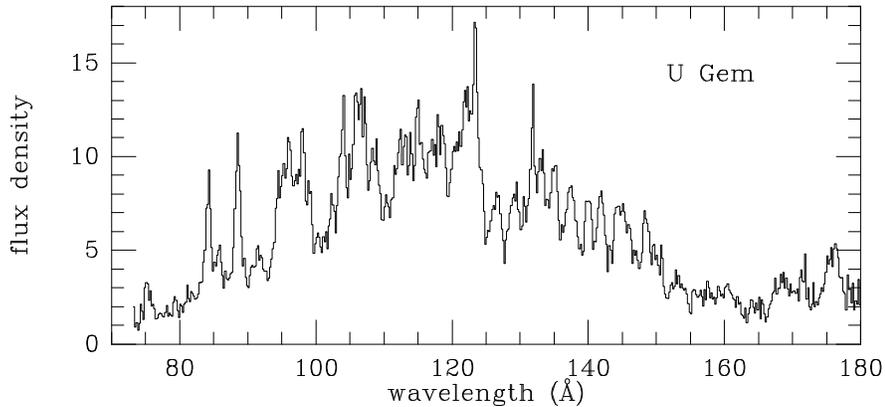}
\caption{{\it EUVE\/} spectrum of U Gem in outburst. Units of flux density
are $10^{-12}$ erg $\rm cm^{-2}$ $\rm s^{-1}$ $\rm \AA ^{-1}$.}
\end{figure}      

\begin{figure}[h] 
\epsfbox[ 86 545 431 700]{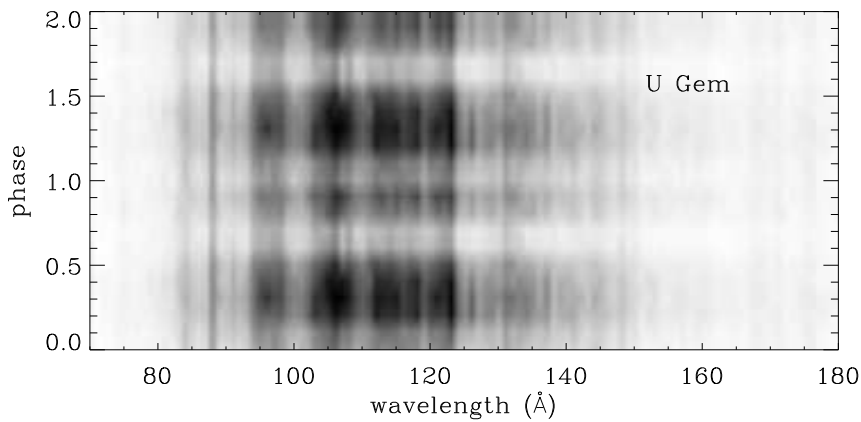}
\caption{{\it EUVE\/} spectra of U Gem in outburst as a function of binary
phase for the ephemeris of Marsh et~al.\ (1990). Note the partial eclipses
at phases $\phi\sim 0.1$ and 0.65 and the persistence through eclipse of
many of the emission lines (e.g., Ne~VIII 2s--2p $\lambda 88.1$).}
\end{figure}      

\subsection{U~Gem}

The {\it EUVE\/} spectrum from the first interval of observations of U Gem in
outburst is shown in Figure~1 (Long et~al.\ 1996). The continuum is reasonably
well described by a blackbody with an effective temperature $T\sim 1.4\times
10^5$~K (12 eV), a hydrogen column density $\NH\sim 3\times 10^{19}~\rm
cm^{-2}$, a luminosity $L\sim 4\times 10^{34}\, (d/90~{\rm pc})^2~\rm
erg~s^{-1}$, and a fractional emitting area $f\equiv L/4\pi\Rwd ^2\sigma
T^4\sim 0.5$. The boundary layer luminosity is $\sim 0.5$ times the accretion
disk luminosity, but U~Gem in outburst would be a weak {\it ROSAT\/} source,
since little flux is emitted shortward of $\sim 80$ \AA \ (0.16~keV).

Superposed on the continuum are emission lines of such species as Ne VI--VIII,
Mg VI--VII, and Fe VII--X; species which dominate in collisionally ionized
gas at $T\sim 3$--$10\times 10^5$~K or in gas photoionized by a blackbody at
$T\sim 2\times 10^5$~K. The observed lines arise from ground-state transitions
with substantial oscillator strengths, and the absence of lines from metastable
levels combined with an estimate for the density of $n_{\rm e}\gax 3\times
10^{11}~\rm cm^{-3}$ argues that the line-emitting gas is relatively cool and
photoionized. That the line-emitting gas is also extended relative to the
continuum is indicated by phase-resolved spectra. Figure~2 shows a gray-scale
representation of the phase dependence of the EUV spectrum of U~Gem. Partial
eclipses occur at phases $\phi\sim 0.1$ and 0.65, but the continuum is much
more severely effected than the lines---note for example the persistence
through eclipse of the Ne~VIII 2s--2p line at 88.1~\AA . These lines could
arise in the wind or in a corona above the accretion disk.

\begin{figure}[t] 
\epsfbox[ 86 545 431 700]{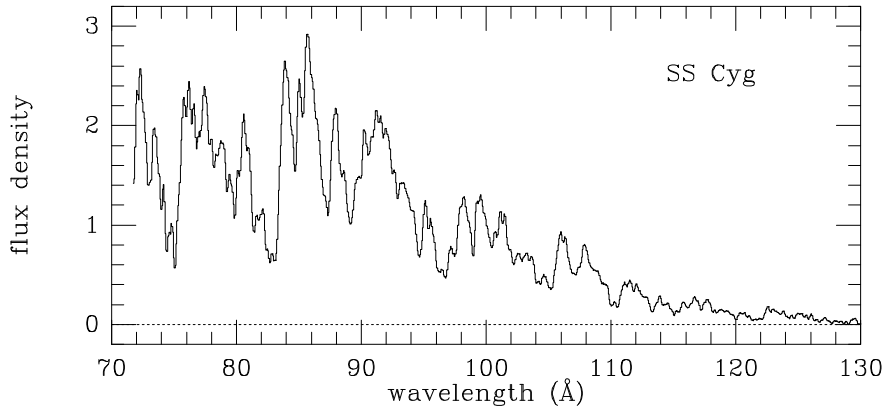}
\caption{{\it EUVE\/} spectrum of SS Cyg in outburst. Units of flux density
are $10^{-12}$ erg $\rm cm^{-2}$ $\rm s^{-1}$ $\rm \AA ^{-1}$.}
\end{figure}      

\subsection{SS~Cyg}

As shown in Figure~3, compared to U~Gem, the {\it EUVE\/} spectrum of SS~Cyg
in outburst is hotter, more absorbed, less luminous, and less obviously
separable into lines and continuum (Mauche, Raymond, \& Mattei 1995). The
spectrum is only crudely approximated by a blackbody with an effective
temperature $T\sim 2.3\times 10^5$~K (20 eV), a hydrogen column density
$\NH\sim 7\times 10^{19}~\rm cm^{-2}$, a luminosity $L\sim 2\times 10^{33}\,
(d/75~{\rm pc})^2~\rm erg~s^{-1}$, and a fractional emitting area $f\sim
3\times 10^{-3}$. The boundary layer luminosity is only $\lax 0.1$ times the
accretion disk luminosity, but the higher effective temperature makes SS~Cyg
a bright {\it ROSAT\/} source. The peaks in the spectrum correspond to
transitions of such species as Ne~VI--VIII, Mg~V--VIII, and Si~IV--VII. These
are roughly the same species as in the spectrum of U~Gem and imply roughly
the same conditions in the line-emitting gas. However, unlike U~Gem, we have
no constraint on the location or physical extent of this gas. Based on the
morphology of the spectrum and the fact that the hydrogen column density is a
factor of $\sim 2$ higher than the interstellar value, the suspicion is that
resonance scattering in the outflowing wind strongly modifies the intrinsic
boundary layer spectrum, but a faithful model of the processes involved is
beyond our current capabilities. To make matters worse, it is not clear that
SS~Cyg possesses a canonical boundary layer: the small effective area, the 
relatively high effective temperature, and the high-amplitude quasi-coherent
oscillations in the EUV/soft X-ray flux suggest that the white dwarf in
SS~Cyg is weakly magnetic ($B\sim 0.1$--1~MG; Mauche 1996a).

\subsection{VW~Hyi}

The comparatively cool and unabsorbed {\it EUVE\/} spectrum of VW~Hyi in
superoutburst is shown in Figure~4 (Mauche 1996b). A blackbody fails miserably
to reproduce the overall spectral distribution, but the effective temperature
must be $T\lax 1.2\times 10^5$~K (10 eV) and the hydrogen column density
must be $\NH\gax 3\times 10^{18}~\rm cm^{-2}$; the bolometric luminosity is
correspondingly uncertain, but the observed 80--420~\AA \ luminosity is
$L\sim 8\times 10^{32}\, (d/65~{\rm pc})^2~\rm erg~s^{-1}$, which is $\sim
0.05$ times the accretion disk luminosity. With little flux emitted shortward
of $\sim 100$~\AA \ (0.12 keV), the boundary layer flux of VW~Hyi in outburst
is unobservable with {\it ROSAT\/}. If similar conditions apply in other
high-$\Mdot $ nonmagnetic CVs, the lack of detections of the soft component
of such systems by {\it Einstein\/} and {\it ROSAT\/} is explained: {\it
a~la\/} Patterson \& Raymond (1985), the boundary layer is simply too cool
to radiate soft X-rays. But, we cannot simply move into the EUV to observe
the boundary layer radiation: if the boundary layer temperature of VW~Hyi were
typical of other nearby CVs, the boundary layer flux would be unobservable
even with {\it EUVE\/}: at $\NH = 3\times 10^{19}~\rm cm^{-2}$, the optical
depth is 1.1 at 100~\AA , 3.2 at 150~\AA , 6.5 at 200~\AA .

\begin{figure}[t] 
\epsfbox[ 86 545 431 700]{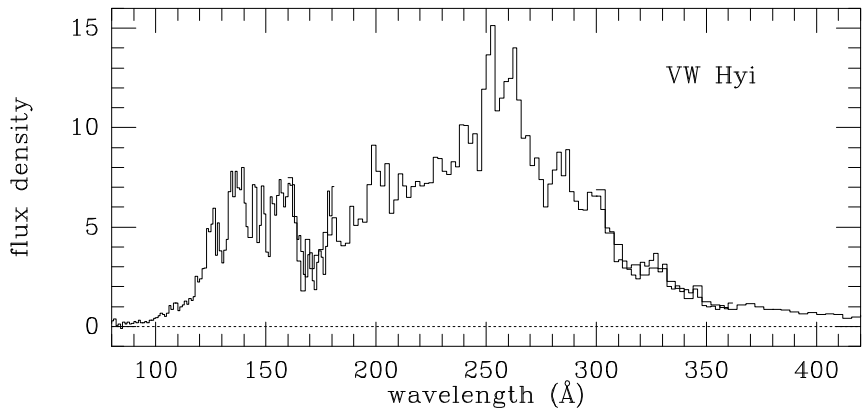}
\caption{{\it EUVE\/} spectrum of VW~Hyi in superoutburst. Units of flux
density are $10^{-12}$ erg $\rm cm^{-2}$ $\rm s^{-1}$ $\rm \AA ^{-1}$.}
\end{figure}      

\section{ASCA Observations}

{\it ASCA\/} has to date observed relatively few nonmagnetic CVs. SS~Cyg was
serendipitously observed in outburst during the PV phase (Nousek et~al.\
1994); VW~Hyi (Mauche \& Raymond 1994) and U~Gem (Szkody et~al.\ 1996) were
observed in quiescence; and observations of about a dozen other CVs have been
obtained, but to date none have been published. {\it ASCA\/} observations of
Z~Cam in outburst were obtained simultaneously with {\it HUT\/} UV and far-UV
spectra in 1995 March. In a preliminary fit performed by M.~Ishida, the SIS
data are satisfactorily modeled by a two-temperature optically thin thermal
plasma spectrum with $kT\sim 0.8$ and 7~keV. Nousek et~al.\ (1994)
fit the {\it ASCA\/} SIS and GIS data for SS~Cyg in outburst with three
components: a two-temperature optically thin thermal plasma spectrum with
$kT\sim 0.8$ and 3.5~keV, plus a thermal brems spectrum with $kT\sim
18$~keV. But, perhaps not surprising, it is now clear and that these two-
and three-temperature fits are not unique descriptions of the data. Kitamura
et~al.\ (1997) fit the SS~Cyg data with a three-temperature optically thin
thermal plasma spectrum with $kT\sim 0.3$, 1, and 8~keV. Done \& Osborne
(1997) fit the same data with the spectrum of an optically thin thermal
plasma with a continuous distribution of temperatures. Both groups now find a
component of the Fe K line at $\sim 6.47$~keV, indicating fluorescence from a
photoionized plasma at $\log\xi\lax 2.4$ and $T\lax 2.5\times 10^5$~K ($kT\lax
22$~eV). This important result implies that the hot gas is in close proximity
to relatively cold material such as the optically thick boundary layer, white
dwarf, and/or accretion disk.

In both Z~Cam and SS~Cyg in outburst, the low-temperature component produces
strong line emission from He-like Si (1.86 keV), S (2.46 keV), and Ar (3.14
keV), but these lines are strongly veiled by the continuum of the
high-temperature component(s). The high-temperature component itself produces
relatively weak lines of H-like Mg (1.47 keV), Si (2.01 keV), and S (2.62
keV), and these lines are themselves veiled by the brems component in the
model of Nousek et~al. The lines which are prominent in the {\it net\/}
spectra are the Fe~K blend at $\sim 6.7$~keV and the Fe~L complex at $\sim
1$~keV. Whereas it is possible with the energy resolution of the {\it ASCA\/}
SIS detector ($E/\Delta E\sim 50$ at 6 keV) to resolve the Fe~K line into
its varies components, nothing can be done with the forest of lines in the
Fe~L complex which dominate the spectrum from $\sim 0.8$ to 1.2~keV.

\begin{figure}[t] 
\epsfbox[ 86 545 431 710]{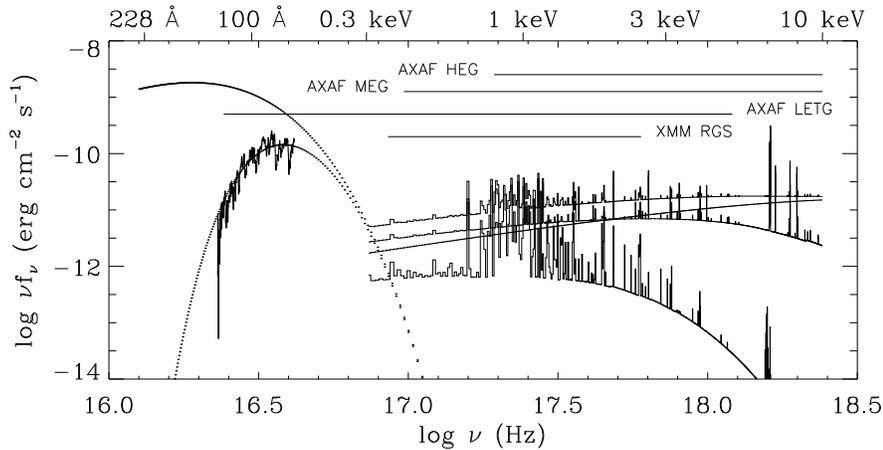}
\caption{Broad-band X-ray spectrum of SS~Cyg in outburst. The horizontal
lines indicate the wavelength/energy ranges of the {\it XMM\/} RGS and
the {\it AXAF\/} LETG, MEG, and HEG.}
\end{figure}     

\section{The Future}

While significant progress in our understanding of the X-ray spectra of
nonmagnetic CVs has and will continue to be made with the growing list of
sources observed by {\it ASCA\/}, it is not too soon to consider the
improvements possible with the next generation of X-ray observatories. To
demonstrate the case for one high-$\Mdot $ CV, a montage broad-band X-ray
spectrum of SS~Cyg in outburst is shown in Figure~5. The figure shows the
{\it EUVE\/} data on the left and the various components of the model of
Nousek et~al.\ for the {\it ASCA\/} data on the right. The dotted curves are
the spectra of a 20~eV blackbody with and without absorption by $\NH = 7\times
10^{19}~\rm cm^{-2}$. The optically thick portion of the boundary layer forms
a (not so) ``big blue bump'' in the EUV/soft X-ray waveband. This component
will be observable (only) with the {\it AXAF\/} LETG with $\sim 5$ times the
effective area and $\sim 5$ times the spectral resolution of the {\it EUVE\/}
SW spectrometer. The Fe L forest is covered by the {\it XMM\/} RGS with
comparable effective area and $\sim 20$ times the spectral resolution of the
{\it ASCA\/} SIS, and by the {\it AXAF\/} gratings with lower effective areas
but $\sim 30$--60 times the spectral resolution. At the Fe K line, the {\it
Astro-E\/} XRS has $\sim 6$ times the effective area and $\sim 10$ times the
spectral resolution of the {\it ASCA\/} SIS. The {\it AXAF\/} LETG will extend
the coverage of the soft component of the boundary layer into the soft X-ray
waveband, supplying both the shape of the continuum to help constrain the
bolometric luminosity, and additional spectral information to constrain the
physical conditions in the optically thick portion of the boundary layer and
the spectral formation processes responsible for its spectrum. The physical
conditions of the optically thin portion of the boundary layer are
constrained by observations with the other instruments. The {\it XMM\/} RGS
and {\it AXAF\/} MEG and HEG will resolve the Fe L forest into its trees.
These same instruments supply the highest spectral resolution for the K-shell
transitions of N through K, while Ca through Fe and beyond are studied best
with the {\it Astro-E\/} XRS. We expect to be able to derive the differential
emission measure of the optically thin gas, its density, and excitation
mechanism (collisional vs.\ photoionization). Where photoionization
applies, the spatial distribution of the gas will be constrained by the
ionization parameter $\xi = L/nR^2$.

\vspace{4mm}

The author is pleased to acknowledge D.~Liedahl and J.~Raymond for kindly
reading and commenting on a draft of this manuscript and M.~Ishida for the
preliminary analysis of the {\it ASCA\/} observation of Z~Cam. This work was
performed under the auspices of the U.S.\ Department of Energy by Lawrence
Livermore National Laboratory under contract No.\ W-7405-Eng-48.

\section{References}

\reference
 Beuermann, K., \& Thomas, H.-C.
 1993, {\it Adv.\ Space Res.\/}, {\bf 13}, \#12, 115.
\reference
 Done, C., \& Osborne, J.~P. 1997, these proceedings.
\reference
 Kitamura, H., Makishima, K., Matsuzaki, K., Ishida, M., \& Fujimoto, R.
 1997, these proceedings.
\reference
 Long, K.~S., Mauche, C.~W., Raymond, J.~C., Szkody, P., \& Mattei, J.~A.
 1996, {\it ApJ\/}, {\bf 469}, 841.
\reference
 Marsh, T.~R., Horne, K., Schlegel, E.~M., Honeycutt, R.~K., \& Kaitchuck,
 R.~H. 1990, {\it ApJ\/}, {\bf 364}, 637.
\reference
 Mauche, C.~W. 1996a, {\it ApJ\/}, {\bf 463}, L87.
\reference
 Mauche, C.~W. 1996b, in Cataclysmic Variables and Related Objects,
 ed.\ A.~Evans \& J.~H.\ Wood (Dordrecht: Kluwer), 243.
\reference
 Mauche, C.~W., \& Raymond, J.~C.
 1994, in New Horizon of X-ray Astronomy --- First Results from ASCA,
 ed.\ F.~Makino \& T.~Ohashi (Tokyo: Universal Academy Press), 399.
\reference
 Mauche, C.~W., Raymond, J.~C., \& Mattei, J.~A.
 1995, {\it ApJ\/}, {\bf 446}, 842.
\reference
 Nousek, J.~A., et~al. 1994, {\it ApJ\/}, {\bf 436}, L19.
\reference
 Patterson, J., \& Raymond, J.~C. 1985, {\it ApJ\/}, {\bf 292}, 535.
\reference
 Ponman, T.~J., et~al. 1995, {\it MNRAS\/}, {\bf 276}, 495.
\reference
 Richman, H.~R. 1996, {\it ApJ\/}, {\bf 462}, 404.
\reference
 Szkody, P., Long, K.~S., Sion, E.~M., \& Raymond, J.~C.
 1996, {\it ApJ\/}, {\bf 469}, 834.
\reference
 van Teeseling, A., Beuermann, K., \& Verbunt, F.
 1996, {\it A\&A\/}, {\bf 315}, 467.
\reference
 van Teeseling, A., Drake, J.~J., Drew, J.~E., Hoare, M.~G., \& Verbunt, F.
 1995, {\it A\&A\/}, {\bf 300}, 808.
\reference
 van Teeseling, A., \& Verbunt, F. 1994, {\it A\&A\/}, {\bf 292}, 519.
\reference
 Wheatley, P.~J., et~al. 1996, {\it A\&A\/}, {\bf 307}, 137.

\end{document}